\documentclass{article}
\usepackage{amsmath}
\usepackage{graphicx}

\input{tcilatex}
\begin{document}

\title{Test of the superluminality of supercurrents induced by a local
electric field in a superconducting-core coaxial cable}
\date{White paper of August 30, 2010}
\author{R. Y. Chiao}
\maketitle

\begin{figure}
\label{coax-figure}
\includegraphics[width=6in]{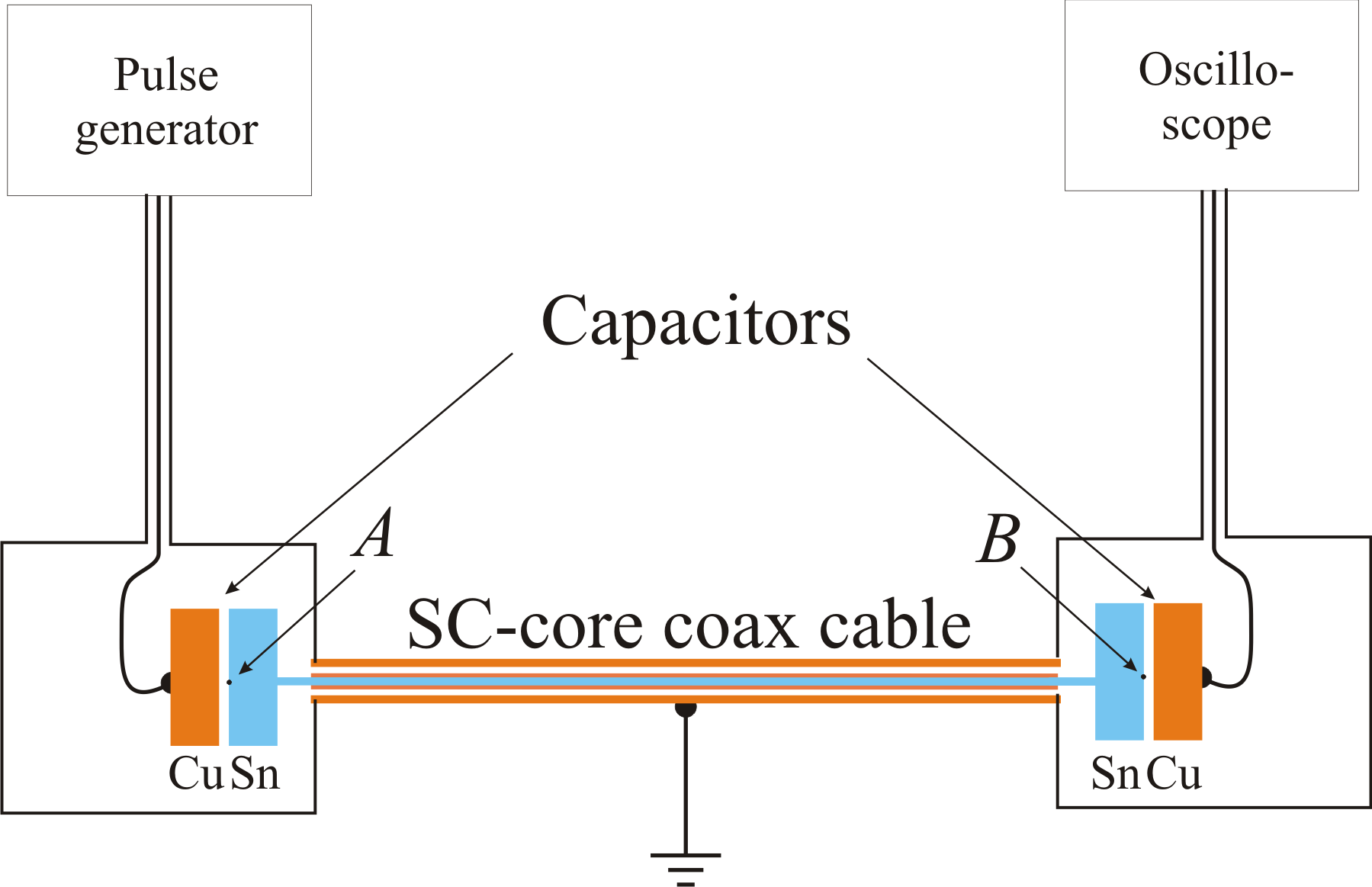}
\caption{Experiment to test the prediction
of superluminal pulse propagation of Cooper pairs in a coax cable with a
superconducting (SC) core. A pulse generator induces pulses of charge inside
a capacitor at point $A$. One plate of the capacitor is a normal metal (copper (Cu)
in $orange$); the other plate is a SC (tin (Sn) in $blue$) and is connected to the SC
core (in $blue$) of a coax cable. Pulses of charge arriving at the capacitor
at point $B$ are detected by a fast oscilloscope.}
\end{figure}

In our Physica E paper \textquotedblleft Do mirrors for gravitational waves
exist?\textquotedblright\ \cite{Prague}, we predicted that superluminal mass
supercurrents will be induced by a gravitational (GR) plane wave which is
incident upon a large, square superconducting (SC) plate. It is the \textit{%
superluminality} of these quantum\ mass supercurrents that leads to the
surprising prediction that the SC plate will act like a mirror which can
reflect the incident GR wave. Specular reflections from SC spheres could
then lead to geometrical cross sections for the scattering of GR radiation.
This is the basis for our proposed Hertz-like experiment, in which pairs of
charged SC spheres with a certain charge-to-mass ratio would efficiently
convert EM radiation into GR radiation, and vice versa. As we continue to
ready the gravitational-wave Hertz-like experiment, it would be helpful to
have firm evidence that superluminal mass currents are indeed possible
within a SC system. Here we propose a simple proof-of-principle experiment
that will allow us to test this basic claim (see Figure 1).

In the proposed experiment, supercurrents are induced by a local electric
field, whereas in the case of GR-wave mirrors they will be induced a GR
wave. Although the interaction between the matter and radiation is more
complex in the GR case, the crucial condition in each case is the production
of superluminal supercurrents by the incident wave. In both cases this
possibility arises from the globally coherent, macroscopic quantum state
occupied by the Cooper pairs, with a well-defined quantum phase established
everywhere inside the body. The resulting globally coherent condensate
wavefunction renders the Cooper pairs fundamentally indistinguishable from
one another within the entire SC system. When a charge pulse whose
characteristic energy is lower than the BCS energy gap of the SC arrives at
point $A$, the nonlocalizability and indistinguishability of the Cooper
pairs imply that the same pulse will be instantaneously (or at least
superluminally) registered at point $B$. For it is fundamentally impossible
to tell, even in principle, whether momentum has been transferred to the
Cooper pairs which are at $A$, or to the Cooper pairs which are at $B$, by
the action of a locally applied electric field.

At the heart of the experiment inside the sample cage of a dilution
refrigerator is a long coaxial cable (with a total length of roughly 5
meters, which will be coiled in order to fit inside the sample cage), with
an inner conductor whose central core is a SC such as tin (indicated in
$blue$). This SC core of the inner conductor is overlaid with a normal metal
sheath such as copper (indicated in $orange$) \cite{Japanese company}, so that
the Cooper pairs can escape from the SC core into this normal sheath, where
they can then break up into normal electrons that propagate along the outer,
normal surface of the inner conductor. The outer conductor of the coax is
also made out a normal metal such as copper (in $orange$). Thus a normal
electromagnetic pulse can propagate \textquotedblleft
luminally,\textquotedblright\ i.e., at the speed of light within the
dielectric of the coax, down the cable. (The dielectric of the cable will be
chosen to be Teflon which has a dielectric constant of around 2.) This
\textquotedblleft luminal\textquotedblright\ pulse will serve as a means of
calibrating the timing measurements for determining the speed of propagation
of the \textquotedblleft superluminal\textquotedblright\ pulses which are
predicted to exist according to our paper \cite{Prague}.

The SC core of the inner conductor is joined (using SC solder joints, for
example) at both ends of the cable to the two SC plates (indicated in $blue$)
of two capacitors (i.e., an input capacitor on the left, and an output
capacitor on the right). The SC plates (in $blue$) are separated from the
normal copper plates (in $orange$) by dielectric spacers in order to form
capacitors. A pulse generator at room temperature produces a train of
nanosecond-scale finite-bandwidth pulses, such as smooth, Gaussian-like
pulses, which propagate down a cryogenic SMA cable into the sample cage
through an SMA\ connector into the interior of a Faraday-cage-like
electronics box containing the input capacitor on the left. A similar \
configuration of a Faraday-cage-like electronics box containing the output
capacitor will be used on the right. Thus a given charge pulse originating
from the pulse generator will be delivered by means of electrostatic
induction to point $A$ of the SC plate of the input capacitor, and the
charge pulse arriving at point $B$ of the SC plate of the output capacitor
will be monitored by means of a fast scope \cite{Sign flip}.

The total charge per pulse induced by the pulse generator near point $A$ of
the input capacitor will be partitioned into two types of charge, viz., type
(i) charge that flows as a normal electrical current along the normal outer
surface of the inner conductor of the coax cable, and type (ii) charge that
flows as a supercurrent within the central SC core of the inner conductor.
Charges of type (i) are normal electrons that result from a pair-breaking
process, in which the Cooper pairs that emerge radially from the central SC
core into the normal sheath surrounding the core of the cable, are broken up
into normal electrons, and end up flowing along the outer surface of the
inner conductor as a normal electrical current. Such charges will be
associated with the \textquotedblleft luminal\textquotedblright\ pulse.
Charges of type (ii) are Cooper pairs that remain unbroken, staying inside
the central SC core and flowing within this core from point $A$ to point $B$
as a supercurrent. Such charges will be associated with the
\textquotedblleft superluminal\textquotedblright\ pulse.

The branching ratio that determines what fraction of the total induced
charge near $A$ ends up after the partitioning process as type (i) charges,
and what fraction ends up as type (ii) charges, will be determined roughly
by the ratio of the capacitance of the cable for an effective length of
cable corresponding the pulse width of a given charge pulse, relative to the
capacitance of the capacitor at $B$ at the end of the cable. For a given
charge pulse of a duration on the order of a nanosecond, which corresponds
to an effective length of cable of roughly 20 cm, this leads to an effective
cable capacitance of 16 picofarads. For an output capacitor with a plate
radius of 1 centimeter, and a gap of 0.3 millimeters, with Teflon as the
dielectric, this leads to a capacitance of 18 picofarads. Therefore, in this
case, we expect the branching ratio to be on the order of unity. Thus we
expect to see a \textquotedblleft double pulse,\textquotedblright\ one
\textquotedblleft luminal\textquotedblright\ and the other \textquotedblleft
superluminal,\textquotedblright\ appearing on a scope trace triggered by the
pulse generator, with approximately equal voltage amplitudes for the two
types of charge pulses.

If the group velocity of the superluminal type (ii) charge pulse is very
much larger than the speed of light \cite{superluminality}, then the
separation in time between the type (i) and type (ii) pulses will be
determined mainly by the time delay of the \textquotedblleft
luminal\textquotedblright\ pulse, which, for a 5 meter long cable, should be
around 24 nanoseconds (assuming an index of 1.4 for Teflon). This should be
easily resolvable using our present oscilloscopes, which have rise times on
the order of 5 nanoseconds. We conclude that the experiment which we propose
here is feasible to perform.

One control experiment would be to warm up the apparatus above the SC
transition temperature, in which case, one expects the \textquotedblleft
superluminal\textquotedblright\ component of the double pulse to disappear,
and only a \textquotedblleft single pulse,\textquotedblright\ viz., only the
\textquotedblleft luminal\textquotedblright\ pulse, to remain. Another
possible control experiment would be to apply a magnetic field that is
larger than the critical field to the cable, in which case, again one
expects only a \textquotedblleft single pulse,\textquotedblright\ viz., only
the \textquotedblleft luminal\textquotedblright\ pulse, to remain. One may
also wish to use two parallel SC-core coax cables connected in parallel to
the input and output capacitor configurations. Then one can ramp the
magnetic field through a hysteresis loop and determine whether the cables
are superconducting or not, by monitoring the presence or absence of trapped
flux inside the SC circuit. This would be an \textit{in situ} method to
establish the presence or absence of superconductivity in our SC samples.

A positive result from this \textquotedblleft
superluminality\textquotedblright\ experiment would imply that not only is
there superluminal transfer of charge occurring from $A$ to $B$, but that
superluminal transfer of mass is also occurring from $A$ to $B$. This is
because each Cooper pair which is being transferred from $A$ to $B$ carries
with it not only charge but also mass. Superluminal mass motions in
superconductors must exist if superluminal charge motions in them were to be
demonstrated to exist. Thus one of the crucial claims of our Physica E paper
would have been successfully demonstrated. On the other hand, if this claim
were to be falsified by experiment, there would be little point in going
ahead with the Hertz-like experiment.

\begin{equation}
\mathbf{j}=\frac{1}{m}\func{Re}\left( \psi ^{\ast }\left\{ \frac{\hbar }{i}%
\nabla -q\mathbf{A}-m\mathbf{h}\right\} \psi \right) \text{ }
\label{prob. curr. density j}
\end{equation}%
where $\psi $ is the condensate wavefunction, $q$ is the charge of a Cooper
pair, $m$ is its mass, $\mathbf{A}$ is the EM vector potential, and $\mathbf{%
h}$ is DeWitt's gravitational vector potential. This is a generalization of
the London equation $\mathbf{j}=\Lambda \mathbf{A}$ for superconductors, in
order to include supercurrents induced by gravitational fields. We also
showed\ that the velocity of the Cooper pairs is given by%
\begin{equation}
\mathbf{v}=\frac{\mathbf{j}}{\psi ^{\ast }\psi }  \label{velocity field}
\end{equation}%
In the special case where there is the absence of radiation fields, and
where the cubic nonlinearity in the G-L equation is negligible, there exists
a plane wave solution of the form $\psi =\sqrt{n}\exp (i\mathbf{k}\cdot 
\mathbf{r})$ of this equation. Then the probability current density given in
(\ref{prob. curr. density j}) becomes%
\begin{equation*}
\mathbf{j}=\frac{\hbar }{2mi}(\psi ^{\ast }(i\mathbf{k})\psi -\psi (-i%
\mathbf{k})\psi ^{\ast })=n\frac{\hbar }{m}\mathbf{k}=n\mathbf{v}\text{ .}
\end{equation*}%
Thus the speed $v$ associated with the current density $j$ is%
\begin{equation*}
v=\frac{\hbar }{m}k\text{ .}
\end{equation*}%
Now, the dispersion relation for de-Broglie matter waves is given by%
\begin{equation*}
\omega =\frac{\hbar k^{2}}{2m}\text{ ,}
\end{equation*}%
which leads to%
\begin{align*}
v_{\text{group}}& \equiv \frac{d\omega }{dk}=\frac{\hbar }{m}k \\
v_{\text{phase}}& \equiv \frac{\omega }{k}=\frac{1}{2}\frac{\hbar }{m}k\text{
.}
\end{align*}%
Hence the velocity $\mathbf{v}$ given in (\ref{velocity field}), which is
associated with the probability current density $\mathbf{j}$ given in (\ref%
{prob. curr. density j}), is the \emph{group} velocity, and not the \emph{%
phase} velocity, of a Cooper pair. The physical meaning of $\mathbf{j}$ is
that it is the quantum transport current density of Cooper pairs. Such
currents transport both the charge $q$ and the mass $m$ of the Cooper pairs
within the SC. In general, the non-relativistic quantum mechanics of
macroscopically coherent wavefunctions of superconductors can lead to
superluminal group velocities, but the front velocity, which is responsible
for relativistic causality, can never exceed $c$. Use of the relativistic
Dirac equation instead of the non-relativistic Schr\"{o}dinger equation does
not change this conclusion, since the Wigner time for the Dirac equation has
also been shown to lead to superluminal group velocities, for example, in
the case of electron tunneling. Gaussian wave packet solutions\ with
superluminal group velocities for both the Dirac and Schr\"{o}dinger
equations lead to the superluminal transport of a quantum particle (such as
the electron in tunneling), so that superluminal mass currents must be
possible in general. In the specific case of superluminal supercurrents
induced by incident gravitational radiation predicted in \cite{Prague}, we
quote from page 246, section 7:

\begin{quotation}
`...the group velocity of a Cooper pair given by (69) is predicted to be
superluminal, even for extremely small values of the dimensionless strain $%
h_{+}$ of an incident GR wave [33]. Using (71), (73), and (78) to solve for $%
\left\vert \mathbf{v}/c\right\vert $, one finds that%
\begin{equation*}
\left\vert \frac{\mathbf{v}}{c}\right\vert =\frac{1}{c}\frac{\Xi }{\Xi -1}%
\left\vert \mathbf{h}\right\vert =\frac{1}{2}\frac{\Xi }{\Xi -1}\left\vert
h_{+}\right\vert \text{ . \ \ \ \ (101)}
\end{equation*}%
Even for an arbitrarily chosen, extremely small value of $\left\vert
h_{+}\right\vert \approx 10^{-40}$ (which, for a 6 GHz GR wave, corresponds
to an incident power flux on the order of $10^{-16}$ W m$^{-2}$), the value
given in (96) leads to a velocity roughly one hundred times the speed of
light. This apparent violation of special relativity suggests that the
response of a superconductor to a GR-wave field will in general be
nonlinear, invalidating our assumption of linearity in (75)\textbf{.}

`However, group velocities much larger than $c$ (infinite, even) have been
experimentally demonstrated [39]. In particular, photon tunneling-time
measurements confirm the \textquotedblleft Wigner\textquotedblright\
transfer time, which is a measure of an effective group velocity broadly
applicable to all quantum scattering processes. Wigner's analysis [40]
assumes a \emph{linear }relation between the initial and final states of a
quantum system, and yields a transfer time that is proportional to the
derivative of the phase of the system's transfer function with respect to
the energy of the incident particle. In the present context, this implies
that the Wigner time will be zero, since the phase of the Cooper-pair
condensate remains constant everywhere, and stays unchanged with time and
energy, due to first-order time-dependent perturbation theory (i.e.,
assuming that no pair-breaking or any other quantum excitation is allowed
[15]). Returning to Figure 1, the Wigner time implies that an observer
located at the center of mass of the superconductor who spots a Cooper pair
at point B during the passage of the wave will see the pair disappear and
then \emph{instantaneously }re-appear at point A. This kind of \emph{%
simultaneity} (as seen by the observer at the center of mass of the system)\
is a remarkable consequence of quantum theory, but it does not violate
special relativity, nor does it invalidate the assumption of linearity.'
\end{quotation}


\begin{thebibliography}{9}
\bibitem{Prague} S.J. Minter, K. Wegter-McNelly, R.Y. Chiao, Physica E 
\textbf{42} (2010) 234--255.

\bibitem{Japanese company} This type of SC coax cable is available from a
Japanese company,.\textquotedblleft Coax Co., Ltd.\textquotedblright

\bibitem{Sign flip} Note that in both the superluminal and luminal cases, a
positively charged pulse at $A$ must appear as a flipped, negatively charged
pulse at $B$. This follows from the conservation of charge.

\bibitem{superluminality} Starting from the Ginzburg-Landau (G-L) theory of
superconductivity, and using the DeWitt minimal coupling rule, we showed
that the supercurrent density induced by electromagnetic and gravitational
fields is given by the expression%
\end{thebibliography}
\end{document}